\documentclass{emulateapj}
\usepackage{natbib}
\usepackage{amsmath}
\usepackage{float}
\usepackage{subfigure}
\usepackage{xcolor}
\usepackage{mathrsfs}
\usepackage{hyperref}

\def\mhcm{\rm m_{H}~cm^{-3}}

\def\kms{\rm km~s^{-1}}
\def\be{\begin{equation}}
\def\ee{\end{equation}}
\catcode`\@=11 

\shorttitle{Horizontal Galactic Wind and Implication on Bubbles' Age}
\shortauthors{Mou et al} 

\begin{document}
\bibliographystyle{apj}
\title
{The Bending Feature of the {\it Fermi} Bubbles: A Presumed Horizontal Galactic wind and Its Implication on the Bubbles' Age}

\author {Guobin Mou \altaffilmark{1,2}, Dongze Sun\altaffilmark{3}, Fuguo Xie \altaffilmark{2}}
 \altaffiltext{1}{School of Physics and Technology, Wuhan University, Wuhan 430072, China; gbmou@escience.cn}
 \altaffiltext{2}{Key Laboratory for Research in Galaxies and Cosmology, Shanghai Astronomical Observatory, Chinese Academy of Sciences, 80 Nandan Road, Shanghai 200030, China}
 \altaffiltext{3}{HongYi Honor College of Wuhan University, Wuhan 430072, China}

\begin{abstract}
There are two spectacular structures in our Milky Way: the {\it Fermi} bubbles in gamma-ray observations and the North Polar Spur (NPS) structure in X-ray observations. Because of their morphological similarities, they may share the same origin, i.e., related to the past activity of Galactic center (GC). 
Besides, those structures show significant bending feature toward the west in Galactic coordinates. This inspires us to consider the possibility that the bending may be caused by a presumed global horizontal galactic wind (HGW) blowing from the east to the west. Under this assumption, we adopt a toy shock expansion model to understand two observational features: (1) the relative thickness of the NPS; (2) the bending of the {\it Fermi} bubbles and NPS. In this model, the contact discontinuity (CD) marks the boundary of the {\it Fermi} bubbles, and the shocked interstellar medium (ISM) marks the NPS X-ray structure. We find that the Mach number of the forward shock in the east is $\sim 1.9-2.3$, and the velocity of the HGW is $\sim 0.7-0.9 c_{\rm s}$. Depending on the temperature of the pre-shock ISM, the velocity of the expanding NPS in Galactic coordinates is around $180-290~\kms$, and the HGW is $\sim 110-190~\kms$.  
We argue that, the age of the NPS and the {\it Fermi} bubbles is about 18--34 Myr.
This is a novel method, independent of injection theories and radiative mechanisms, for the estimation on the age of the {\it Fermi} bubble/NPS. 
 \end{abstract}


\section{INTRODUCTION}

Two giant gamma-ray bubbles located symmetrically above and below the Galactic plane are revealed by the {\it Fermi} Gamma-ray Space Telescope \citep{Dobler2010, Su2010, Ackermann2014}. 
 Each of the bubbles is roughly $\sim 50^{\circ}$ in height, and $\sim 40^{\circ}$ in width. 
If the Sun-GC distance is 8.5 kpc \citep{Ghez2008}, considering the projection effect of a three-dimensional structure, the physical height is 9-10 kpc, while the width is 5-6 kpc (e.g., \citealt{Guo2, Mou2015, Sarkar2018}). 
More interestingly, in X-ray band, {\it ROSAT} All-Sky Survey at 1.5 keV showed a giant limb-brightened feature in the northeast sky which is usually called the North Polar Spur (NPS), and a less significant counterpart in the northwest sky which is usually called the NPS-W  (\citealt{Snowden1997}). The distance of NPS is still under debate: it may be located at Galactic halo with distance of several kpc (e.g., \citealt{Sofue2000, Kataoka2013, Sun2014, Sofue2015, Sofue2016, Lallement2016, Akita2018}), or may be a local structure near the sun with a distance of several hundred pc (e.g., \citealt{Egger1995, Wolleben2007, Puspitarini2014}). However, as shown later in Figure \ref{plot1}, morphologically the NPS structure seems to be located just outside and surrounding the north {\it Fermi} bubble, which implies that they may share a common origin \citep{Su2010}. 

The formation and radiative mechanism of the {\it Fermi} bubbles and the NPS structure are still under debate. These structures may either relate to wind driven by star formation in the Galactic center (GC) \citep{Crocker2011, Crocker2014, Crocker2015, Carretti2013, Sarkar2015}, or relate to the past activities of supermassive black hole residing in the GC --- Sgr A* (\citealt{Guo1, Guo2}; \citealt{Yang2012, Yang2013, Yang2017}; \citealt{Zubovas2012}; \citealt{Mou2014, Mou2015}). In those Sgr A*-induced scenarios, it is assumed that Sgr A* has experienced a much more luminous active period (compared to the current dim non-active period) that began millions of years ago (see the review by \citealt{Totani2006}).
 During that active period, a huge amount of energy has been ejected out, through either collimated jet or un-collimated wind, from the accretion system of Sgr A*. The high speed outflowing material rushes into the interstellar medium (ISM) in the galactic halo, creating expanding shock structures. In these shock-based scenarios, geometrically the {\it Fermi} bubble is bounded by the contact discontinuity (CD), while the NPS structure corresponds to the post-shock ISM (cf. Sec.\ \ref{ToyModel} and the shaded region in the bottom panel of Figure \ref{plot2} below).

\begin{table*}
\begin{center}
\centerline{Table 1 -- Theoretical models for the {\it Fermi} bubbles}
\vspace{0.2 cm}	
\begin{tabular}{c c c c c}	
\hline
 Dynamical Model     & Injected Outflow Velocity  & Total Injected Energy & Radiation Model   & $\tau_{\it Fermi}$ (Myr) \\	
\hline
Jet \footnote{References: \citealt{Guo1, Guo2}}  & $V_{\rm jet}=0.1\ $c & $\sim 10^{56-57}\ $erg &  leptonic          & 1--2 \\
Jet \footnote{Ref: \citealt{Yang2012, Yang2013, Yang2017}}  & $V_{\rm jet}=0.025\ $c & $\sim 10^{57}\ $erg  &  leptonic          & $\sim\ $1    \\
Quasar wind  \footnote{Ref: \citealt{Zubovas2011, Zubovas2012}}        & $V_{\rm wind}=0.1\ $c & $\sim 10^{57}\ $erg &  leptonic          & $\sim\ $6    \\
Hot accretion wind \footnote{Ref: \citealt{Mou2014, Mou2015}} & $V_{\rm wind}\simeq 0.05\ $c & $\sim 10^{55-56}\ $erg & hadronic dominated  & 7--12\\
Star-formation-driven wind \footnote{Ref: \citealt{Crocker2011, Crocker2014, Crocker2015}} & $V_{\rm wind}\simeq 1000\kms $ & $\sim 10^{55-57} $erg & hadronic  & 200--$10^{3}$\\
Star-formation-driven wind \footnote{Ref: \citealt{Sarkar2015,Sarkar2018}} & $V_{\rm wind} \simeq 1000\kms$ & $ \sim 10^{55}$ erg & leptonic & 30 \\
\hline
\label{tab1}																						
\end{tabular}
\end{center}
\end{table*}

Apart from the understanding on the morphological structures, there are two possible models existing for the $\gamma$-ray emission of the {\it Fermi} bubbles, i.e., the leptonic model and the hadronic model. The resulting age estimations on the {\it Fermi} bubbles are highly different, which are summarized in Table \ref{tab1}. In the leptonic model, the gamma-ray photons come from inverse Compton (IC) scattering on soft photons (including starlight and cosmic microwave background) by cosmic ray electrons (CRe). Since the cooling timescale of CRe at energies $\sim$1 TeV that demanded in fitting the $\gamma$-ray SED is shorter than $\sim$1 Myr (Figure 28 in \citealt{Su2010}), the age of the {\it Fermi} bubbles should also be shorter than this timescale, which demanding a quite strong power of the jet \citep{Guo1, Guo2,Yang2012, Yang2013, Yang2017}.

In the hadronic model, on the other hand, the gamma-ray photons are generated during the $\pi^{0}$ decays which are produced in collisions between the CR protons and thermal nuclei (so-called $pp$ collisions). The cooling timescale of CRp is $t_{pp}\ga 3.5\times 10^{9}~{\rm yr}~(0.01~{\rm cm^{-3}} n^{-1}_{\rm H})$ (\citealt{Crocker2014}). Therefore the hadronic model predicts a much longer age of the {\it Fermi} bubbles than the leptonic model, e.g., it is around 7--12 Myr in \citet{Mou2014, Mou2015}, or a few hundreds Myr in \citet{Crocker2014, Crocker2015}.

In this work, we focus on the relative thickness of NPS and the bending feature of both the {\it Fermi} bubbles and the NPS. The former feature implies the strength of forward shock, while the latter may implies the existence of a presumed horizontal galactic wind (HGW). Based on this assumption, we built a toy model to estimate the HGW's velocity, the expanding velocity of NPS (east side), and the age of the {\it Fermi} bubbles.
In Sec.\ \ref{model} we present the observations of the {\it Fermi} bubbles and the NPS. Then we present our model, including its key assumptions and the corresponding calculations. This provides a novel method, independent of injection theories (i.e., accretion-driven jet, accretion-driven wind, or star-formation-driven wind) and radiative mechanisms (leptonic or hadronic) for the age estimation. We briefly discuss our results in Sec.\ \ref{discussion}.

\section{Our Model and Results}\label{model}
\subsection{Morphological Information of the NPS and the {\it Fermi} Bubbles} \label{Morphology}

\begin{figure}
 \centering
 \includegraphics[width=7.5cm]{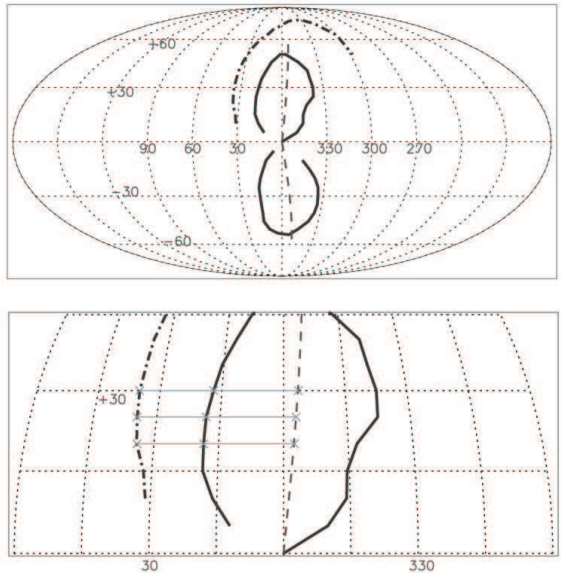}\\
 \caption{ \emph{Upper panel}: morphologies of the {\it Fermi} bubbles (solid curve, see \citealt{Su2010}), and the NPS structure in 1.5 keV band in Galactic coordinates (see \citealt{Snowden1997}). The dot-dashed curve masks the outer boundary of the NPS. Both of the {\it Fermi} bubbles and NPS seem bending toward the west significantly. The north {\it Fermi} bubble show a bending angle of $\theta_{\rm bend}\sim 7^\circ$ (dashed curve), while the south bubble shows a bending angle of $\sim 11^\circ$ (dashed curve). \emph{Bottom panel}: zooming in of the middle part ($b=0-45^\circ$, $l=60^\circ-0-300^\circ$) of the \emph{Upper} panel. We show three gray solid lines to measure the thickness of the NPS, and the distance of the outer edge of the NPS to the expansion center. The gray crosses mark the measuring points: expansion center, east edge of NPS and NFB. }
 \vspace{-0.1cm}
 \label{plot1}
\end{figure}

In Figure \ref{plot1} we plot the morphologies of the NPS in 1.5 keV band (\citealt{Snowden1997}) and the {\it Fermi} bubbles (solid curve,\citealt{Su2010}). Evidently both the NPS and the {\it Fermi} bubbles seem to bend toward the west. 
The bending angle is about $\theta_{\rm bend}\approx 6-8^\circ$ for the north {\it Fermi} bubble (NFB).
Besides, the X-ray observations also indicate that the NPS-W is fainter than the NPS (not shown here, refer to \citealt{Snowden1997}). The asymmetric morphologies imply that, they probably have suffered from a presumed HGW in the halo blowing from the east to the west.

Another result from this figure is that, we can measure ratio between the thickness of the NPS ($\Delta R_{\rm NPS}$) and the distance from the outer boundary of NPS structure to the expansion center ($d_{\rm NPS}$). 
Such information will later be used to determine the Mach number of the forward shock.
From the zoom-in plot (bottom panel of Figure \ref{plot1}), we can estimate the distance between the east edge of the NPS or NFB, and the expansion center. The formula is
\be
d_{\rm NPS, NFB}=\frac{d_{\odot} \sin(l_1-l_2)}{\cos(b) \cos(l_2)}
\ee
in which $d_{\odot}=8.5$ kpc, $b$ is the latitude of measuring points, $l_1$ and $l_2$ are the longitude of the east edge of NPS/NFB and the expansion center (EC), respectively (see gray crosses plotted in the bottom panel of Figure 1).
Our results are shown in Table 2. 

\begin{table*}
\begin{center}
\centerline{Table 2 -- Results of Measurement on NPS and the North Fermi Bubble (NFB) }
\vspace{0.2 cm}	
\begin{tabular}{c c c c c c c}	
\hline
 b & $l_1$ -(Long. of NPS) & $l_1$-(Long. of NFB) & $l_2$ - Long. of EC  \footnote{Longitude of Expansion Center, when the bending angle equals 6$^\circ$, 7$^\circ$, or 8$^\circ$} & $d_{\rm NPS}$ \footnote{distance from east edge of NPS to expansion center, when the bending angle equals 6$^\circ$, 7$^\circ$, or 8$^\circ$} & $d_{\rm NFB} \footnote{distance from east edge of NFB to expansion center,}$ & 
$f_1$ \footnote{relative thickness: $(d_{\rm NPS}-d_{\rm NFB})/d_{\rm NPS}.$ } \\	
\hline
20$^\circ$ & 33.9$^\circ$ & 18.5$^\circ$ & -2.1$^\circ$/-2.5$^\circ$/-2.9$^\circ$   & 5.32/5.37/5.43 & 3.18/3.24/3.30 &  0.401/0.396/0.391  \\
25$^\circ$ & 34.7$^\circ$ & 18.3$^\circ$ & -2.6$^\circ$/-3.0$^\circ$/-3.4$^\circ$  & 5.69/5.74/5.80 & 3.35/3.41/3.47 & 0.411/0.406/0.401 \\
30$^\circ$ & 35.0$^\circ$ & 17.0$^\circ$ & -3.1$^\circ$/-3.6$^\circ$/-4.1$^\circ$ & 6.06/6.13/6.21 & 3.38/3.46/3.54 & 0.443/0.436/0.429 \\
\hline
\label{tab2}																						
\end{tabular}
\end{center}
\end{table*}

\subsection{Our Toy Model} \label{ToyModel}

Our toy model is based on three assumptions. Firstly, the {\it Fermi} bubbles and the NPS share a common origin, related to the past activities of the GC. Secondly, there exists a global east-to-west blowing HGW in the Galactic halo, which bends the {\it Fermi} bubbles and the NPS to certain degrees toward the west.
Thirdly, we assume that the nucleus activity is a constant over time. We note that if the density of ISM follows the observed form $n_{e} \propto r^{-2.1}$ in \citet{Miller2013}, the velocity of the forward shock virtually does not change with time for a constant injected power, based on analytical solutions of the shock's evolution (cf. Equation 9 in \citealt{Mou2014}). 

\begin{figure}
 \centering
 \includegraphics[width=8cm, angle=0]{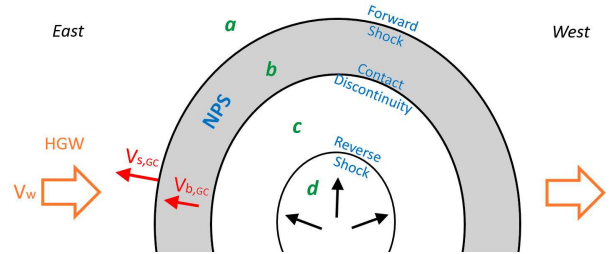}
 \caption{The toy shock model for the {\it Fermi} bubble and NPS structures. Supersonic outflow from the GC interacts with the ambient ISM, forming shock structures. The space can be divided into four zone: (\emph{a}) pre-shock ISM; (\emph{b}) post-shock ISM; (\emph{c}) post-shock outflow; (\emph{d}) supersonic outflow (\citealt{Weaver1977}). The outer edge of the NPS is the forward shock front, and the {\it Fermi} bubbles edge is the CD. We assume a HGW blowing from the east to the west, as shown by orange arrows. }
 \label{plot2}
\end{figure}

Now we provide more details of our toy model. The schematic diagram is shown in Figure \ref{plot2}. The supersonic outflow is injected from the GC. The outflow may be in the form of star-formation-driven wind in the central regions of our galaxy, jet, wind from a quasar state accretion disk, or wind from hot accretion disk (see \citealt{YuanNarayan2014, Yuan2015} for hot accretion wind). The exact form of the outflow is irrelevant for the investigations here. As shown in Figure \ref{plot2}, shock structures are generated by the supersonic outflow. The ram pressure of the supersonic outflow is balanced by the thermal pressure of the post-shock outflow gas  (region \emph{c}) at the reverse shock front. The interface between the post-shock outflow (region \emph{c}) and the post-shock ISM (region \emph{b}) is called the CD.

The boundary of the {\it Fermi} bubbles is generally determined by the CD, because the magnetic field in the post-shock ISM is parallel to the CD, which can bound the CRs inside the CD \citep{Yang2012}, or outside but close to the CD \citep{Mou2015}. 
According to simulations of jet/quasar wind/hot accretion wind models, cross the CD from region \emph{c} to region \emph{b}, there is a sharp increase in density, and a sharp decrease in temperature.
The forward shock front compresses the ISM, making a hotter and denser post-shock ISM (region \emph{b}) compared to the pre-shock ISM (region \emph{a}). Thus the shocked ISM (region \emph{b}) will be the brightest in X-ray band among regions \emph{a-c}, likely corresponding to the NPS structure, that locates just surrounding the north {\it Fermi} bubble \citep{Su2010}. Such a scenario is adopted in almost all the models mentioned above.

If there is an ``east-to-west" blowing HGW, both the {\it Fermi} bubbles and the NPS bend toward the west. Besides, the Mach number of the eastern forward shock will be larger than that of the western counterpart. Consequently, the density and temperature in the east post-shock ISM would be higher than those in the west post-shock ISM. Consequently, the NPS in the east side will be brighter than the NPS-W in X-ray image, a phenomenon already observed. A reliable measurement of the thickness of NPS can be made in the east side, because of its clear image.

The rationality of the existence of the HGW may be controversial. In addition to two aspects mentioned above, we here add some more arguments. First, according to ROSAT X-ray observations, in north galactic hemisphere, the outflows close galactic center is perpendicular to the galactic plane (Fig 19. in \citet{Su2010}, also see \citet{Zubovas2012} for simulations on interactions between quasar outflows and massive Central Molecular Zone in GC). However the NPS/Fermi bubble caused by outflow is significantly bending toward the west in higher latitude. 
Therefore, we believe that the bending is most likely due to HGW, not for other reasons.
Second, head-tail radio galaxies are very common, in which jets appear to be bent by ram pressure of winds. Those wind velocities range from a few hundred to several thousand kilometers per second (e.g., see Table 5 in \citealt{OEO1993}). Besides, in our local group, M31 (l=121$^{\circ}$, b=-22$^{\circ}$, roughly east direction in Galactic coordinates) is approaching us at 110 $\kms$. Hence galaxies are not stationary in the group, and proper motions of the galaxies and gas can more or less produce winds. Therefore we think that, the existence of such a kind of wind is not so unreasonable.

\subsection{The Relative Thickness of NPS}
With subscript ``1'' representing the \emph{east} part, the shock velocity in the frame of pre-shock ISM, i.e., $V_{\rm s,1}$, can be written as,
\be
V_{\rm s,1}=V_{\rm s,GC}+V_{\rm w}=M_{1}c_{\rm s},  \label{Vs1} \\
\ee
and the velocity of CD in this frame is,
\be
V_{\rm cd,1}=V_{\rm cd,GC}+V_{\rm w},
\ee
where $V_{\rm s,GC}$ is the velocity of the forward shock in GC frame, $V_{\rm w}$ is the HGW velocity in GC frame, $c_{\rm s}$ is the adiabatic sound speed of the pre-shock ISM, and $M_{1}$ is the Mach number of the forward shock, $V_{\rm cd,GC}$ is the velocities of the contact-discontinuity in the frame of GC.

We firstly use the ratio between the thickness of the NPS ($\Delta R_{\rm NPS}$) and the distance from the outer boundary of this structure to the expansion center ($d_{\rm NPS}$), i.e.,
\be
f_1 \equiv \frac{\Delta R_{\rm NPS}}{d_{\rm NPS}} = \frac{V_{\rm s, GC}-V_{\rm cd, GC}}{V_{\rm s, GC}}=\frac{r_{t}(M_1)}{1-V_{\rm w}/(M_{1}c_{\rm s})},  \label{Xthick}
\ee
where $r_{t}(M_1) \equiv (V_{s,1}-V_{cd,1})/V_{s,1}$ is the relative thickness of shocked ISM in HGW's frame, and $r_{t}(M_1)$ is mainly determined by the Mach number $M_{1}$ (see below for the relationship). 
The range of $f_1$ depends on the bending angle and measuring points, and ranges from 0.391 to 0.443 (Table 2).
The sound speed of unshocked gas in galactic halo $c_{s}$ is determined by observations, in which the temperature of galactic halo is estimated to be $T=1-2 \times 10^{6}$ K (\citealt{Kuntz2000, Miller2013, Miller2016, Kataoka2018}). 
Therefore the sound speed $c_s$ is 150-210 $\kms$ since $c^2_s=\gamma k_{B}T/(\mu m_{H})$ in which $\mu=0.63$ for solar abundances.

\begin{figure}
 \centering
 \includegraphics[width=8cm, angle=0]{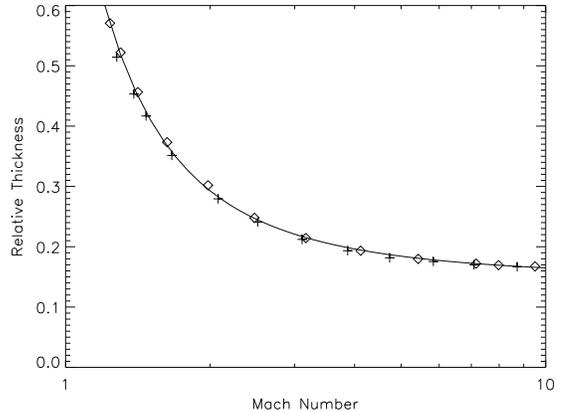}
 \caption{Relationship between forward shock's Mach number $M_f$ and relative thickness $r_t$. Crosses mark the simulation results with $\rho_{\rm out}(r_{in})=0.1 \mhcm$, diamonds mark the results with $\rho_{\rm out}(r_{in})=10.0 \mhcm$ ($r_{\rm in}$=0.1 kpc). The solid line shows the fitting curve (Equation \ref{fitting}). Although there are two orders of magnitude difference in outflow densities, the $M_f - r_t$ relationships in both cases are highly consistent.}
 \label{plot3}
\end{figure}

We have made numerical simulations to find the relationship between $M_f$ and $r_t$ ($M_f$ is the forward shock's Mach number and $r_t$ is the relative thickness defined as the ratio between the thickness of shocked ISM and forward shock's radius). We use ZEUSMP code 
(\citealt{Stone1992, Hayes2006}), and choose 2.5 dimensional Spherical coordinate by assuming that it is symmetric in $\phi$-direction. The scale of $r$-direction is set to be 0.1--10 kpc, and it is divided into 900 uniform grids with $dr(i+1)/dr(i)=1.005$. The initial ISM is assumed to be isothermal, and follow a $\rho_{\rm ISM} \propto r^{-2}$ law while the gravitational potential is $\Phi=-2\sigma^{2}/r$ ($\sigma=124 ~\kms$). We inject an isotropic outflow of a certain power at the inner boundary each time, and analyze the evolution of the shock structures. For the initial density obeying $r^{-2}$ law, we can obtain shock structures evoluting linearly with time (i.e., the velocities of forward shock, CD remain constant over time). Our results are plotted in Figure \ref{plot3}, in which Mach numbers cover a range of 1.24--10. We fit the $M_f - r_t$ relationship with an analytical expression:
\be
3/M_{f}=\ln(r_{t}-0.11)+3.2  ~. \label{fitting} \\  
\ee
In this work, $M_f$ is $M_1$ mentioned above. According to the $M_f - r_t$ relationship, if there is no HGW or $V_{\rm w}/(M_1 c_s)$ is tiny, 
$M_1$ is $1.43 \sim 1.55$ since $r_t=f_1$ in this case, while if $V_{\rm w}/(M_1 c_s)$ is not negligible, $M_1$ should be larger than this range. 

\subsection{The Bending of Fermi Bubbles}
We then make use of the bending angle of the north {\it Fermi} bubble. The bending of a jet in a cross wind has been studied for decades \citep{Burns1979, JO1979, OEO1993}, 
and the bending formula is:
 \be
 \rho_{\rm out} \frac{v^2_{\rm out}}{l_{\rm bend}} \sim \rho_{0} \frac{V^2_{\rm w}}{l_{\rm press}} \label{bend}
 \ee
 in which $\rho_{\rm out}$ and $v_{\rm out}$ are the density and velocity of outflows (jets or winds launched from GC in different models), $l_{\rm bend}$ is the radius of curvature, $l_{\rm press}$ is the length scale over which the ram pressure acts, and $\rho_0$ is the density of pre-shock gas in galactic halo. 
Here $l_{\rm bend} \sim H_{\rm FB}/2\theta$, and $l_{\rm press} \sim D_{\rm FB}$ 
( $H_{\rm FB}$ and $D_{\rm FB}$ are height and width of Fermi bubbles respectively).
 Regarding the projection effect of a three-dimensional structure, we consider three physical lengths of NFB based on simulation works:
 $(H_{\rm FB}, D_{\rm FB})$=(9 kpc, 4.5 kpc) in \citet{Sarkar2018}, (9.4 kpc, 5 kpc) in \citet{Mou2015}, and (10 kpc, 6 kpc) in \citet{Guo2}.
 Then $f_2 \equiv l_{\rm bend}/l_{\rm press}=6.0 \sim 9.6$.
Inside the shock structure, the ram pressure of supersonic outflows is comparable to the thermal pressure of shocked ISM: $\rho_{\rm out} v^2_{\rm out}=\Lambda P_{\rm b}$ in which $\Lambda \sim 1$. 

Regarding the Rankine-Hugoniot conditions at the forward shock front in the HGW frame, we have
 \begin{gather}
 \rho_{0} V_{\rm s,1}=\rho_{\rm b,1} (V_{\rm s,1}-V_{\rm b,1})  ,  \label{RHmass} \\
 \frac{1}{\gamma} \rho_{0} c^{2}_{\rm s}+\rho_{0} V^{2}_{\rm s,1}=\frac{1}{\gamma} \rho_{\rm b,1} c^{2}_{\rm s,1}+\rho_{\rm b,1} (V_{\rm s,1}-V_{\rm b,1})^{2}  , \label{RHmomt} \\
 \frac{1}{2} V^{2}_{\rm s,1}+\frac{5c^{2}_{\rm s}}{2\gamma}=\frac{1}{2} (V_{\rm s,1}-V_{\rm b,1})^{2}+\frac{5c^{2}_{\rm s,1}}{2\gamma} . \label{RHcon}
\end{gather}
in which $\rho_{\rm b,1}$ and $c_{\rm s,1}$ are the density and sound speed of the post-shock ISM (\emph{east} part), respectively.
With adiabatic index $\gamma=5/3$, we have $\rho_{b,1}=\rho_0 ~ 4M^2_1/(M^2_1+3)$ and $c^2_{s,1}=c^2_{s}(0.3125M^2_1+0.875-0.1875/M^2_1)$.
Therefore, thermal pressure of post-shock ISM
\be
P_{b}= \rho_{b,1} c^2_{s,1}/\gamma=\rho_{0}c^2_s \frac{3.75M^4_1+10.5M^2_1-2.25}{5M^2_1+15}. \label{Pb}
\ee
With Equation \ref{bend} and \ref{Pb}, we finally obtain:
\be
V^2_{\rm w}=c^2_s \frac{3.75M^4_1+10.5M^2_1-2.25}{f_2(5M^2_1+15)} ~. \label{Bending}
\ee

\subsection{Results}

Gathering the equations \ref{Xthick}, \ref{fitting}, and \ref{Bending}, we solve the equations in three cases, respectively:

1) when the bending angle equals 6$^{\circ}$, the range of $f_1$ is 0.401-0.443, and $f_2$ is 8.0-9.6,

2) when the bending angle equals 7$^{\circ}$, the range of $f_1$ is 0.396-0.436, and $f_2$ is 6.8-8.2,

3) when the bending angle equals 8$^{\circ}$, the range of $f_1$ is 0.391-0.429, and $f_2$ is 6.0-8.0.

\begin{figure}
 \centering
 \includegraphics[width=8cm, angle=0]{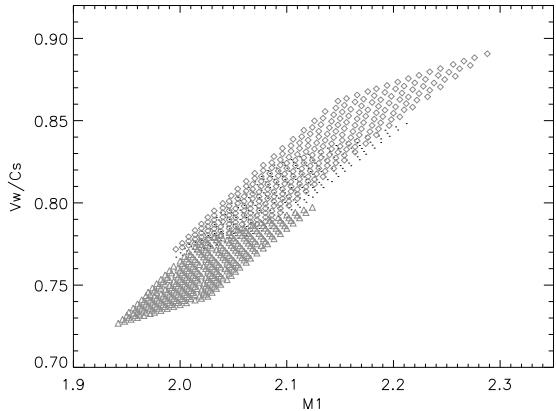}
 \caption{Solutions of $V_{\rm w}$ and $c_s$ according to the range of $f_1$ and $f_2$ in the cases of a bending angle that equals 6$^{\circ}$ (gray triangles), 7$^{\circ}$ (black dots), and 8$^{\circ}$ (gray diamonds). }
 \label{plot4}
\end{figure}

The solutions are shown in Figure \ref{plot4}, and $M_1=1.94-2.29$, $V_{\rm w}/c_s=0.73-0.89$. 

Considering the sound speed of halo $c_s\sim 150-210~\kms$ ($T=1-2\times 10^{6}$ K), we conclude that:

(1) the the velocity of HGW is $110-190~\kms$,  

(2) the forward shock of NPS (\emph{east} part) in the frame of GC moves with $V_{\rm s,GC}=(M_1-V_{\rm w}/c_s) c_s=180-290 ~\kms$. 

We then estimate the age of NPS by $\tau \sim d_{\rm NPS}/V_{\rm s,GC}=18-34$ Myr.
This is also the age of the  {\it Fermi} bubbles.

\section{SUMMARY AND DISCUSSION}\label{discussion}
In this Letter, we estimate the age of {\it Fermi} bubbles/NPS, based on two observational facts: 1) after the shock expansion, the NPS shows a measurable thickness, which is due to the difference in the velocities of the forward shock front and the CD; 2) both the {\it Fermi} bubbles and the NPS are asymmetric, with a bending angle $\theta_{\rm bend}\approx 6-8^\circ$. 
We make an assumption that the bending may be caused by a presumed HGW blowing from east to west in Galactic coordinates, and we obtain the forward shock velocity of NPS (\emph{east}) in Galactic coordinates and velocity of HGW. 
We find that the HGW's velocity
is significantly lower than the unreasonably large value of 750 km/s presented in \citet{Yang2012}, which may eventually strip out the gas of the Milky Way. The main reason is that, in their jet model, the forward shock is so strong that a very fast HGW is demanded to bend the bubble to the observed angle. 

The existence of the HGW can be proved/disproved by absorption lines like O VII absorption lines. The doppler shift in wavelength for O VII Kα line (21.6 Angstrom) is 0.008-0.013 A for the velocity of 110-190 $\kms$ in the viewing direction. However, considering the projected effect of velocity in line of sight and contribution of absorbers closer to the galactic plane which suffered less from HGW, the doppler shift is not so noticeable, which may be marginally resolved by X-ray telescope if the spectral quality is good enough (e.g., O VII absorbers for NGC 3783 in the west direction show redshift of 45-128 $\kms$ with 1-$\sigma$ confidence in \citealt{Gupta2012}). 

\subsection{Comparison with Other Observational Results}\label{otherobs}

We here compare our results with some observations.

 1) According to observations, the temperature of NPS is around 0.3 keV (\citealt{Kataoka2013, Kataoka2016}), or 0.4--0.5 keV (\citealt{Miller2016}). 
If the pre-shock gas temperature is $T_{0}=0.1-0.2$ keV, 
the Mach number of 1.94-2.29 in our fiducial model implies that the temperature of NPS at lower latitudes ($b\sim20^{\circ}-30^{\circ}$) is around 0.2-0.46 keV, roughly consistent with observations.   

 2) \citet{FangJiang2014} found that the shock-expansion velocity of the shocked ISM surrounding the {\it Fermi} bubbles is 200--300 $\kms$, roughly consistent with the result in our model.

 3) \citet{Ackermann2014} found that the gamma-ray image of Loop I (mainly coincident with, but more extended than NPS) is just surrounding the north {\it Fermi} bubble and the lower part of the south bubble with a $\gamma$-ray photon index of $\Gamma\approx 2.4$ (see their Figure 13), implying that the power-law distribution index of the corresponding non-thermal electrons or protons is $p_{\rm PL}\approx 2.4$. According to first-order {\it Fermi} acceleration theory, the Mach number of the forward shock is 3--4 (\citealt{Drury1983}), which is also roughly consistent with our result.

 \subsection{Implications on the {\it Fermi} Bubbles' Age} \label{modeldiscussion}

Our model provides a novel method to estimate the age of the {\it Fermi} bubbles/NPS, independent of detailed dynamical and radiative models. The age is $\tau \sim$ 18-34 Myr. This
is roughly consistent with the $\tau_{\it Fermi} = 7-12$ Myr in the ``hot accretion wind'' model (\citealt{Mou2014, Mou2015}), $\tau_{\it Fermi}= 6$ Myr in ``quasar wind'' model (\citealt{Zubovas2011, Zubovas2012}), and $\tau_{\it Fermi} \approx 30$ Myr in ``star-formation-driven wind+leptonic scenario" model (\citealt{Sarkar2015, Sarkar2018}). However, it is much longer than the ages of $\tau_{\it Fermi}\approx1-2$ Myr in the jet model (\citealt{Guo1}; \citealt{Guo2}; \citealt{Yang2012}; \citealt{Yang2013}), and much shorter than $10^{8}-10^{9}$ yr in the ``star-formation-driven wind'' model (\citealt{Crocker2011, Crocker2014}).

Our result favors the ``hot accretion wind'' model, the ``quasar wind'' model, and the ``star-formation driven wind+leptonic scenario" model.
But it is worth noting that the hypothesis that the past nucleus activity does not change over time may have a significant impact on the final result. Numerical simulations under several typical varying activities are worthwhile in the future.

\acknowledgements
We thank the anonymous reviewer's comments for our first draft, and Wei-Xiao Wang, Hui Yang for their help. This project was supported in part by the Natural Science Foundation of China (Grant No. 11703022, 11833007, 11873074, 11622326), and the National Program on Key Research and Development Project (Grants No. 2016YFA0400803). G.B.M is also supported in part by the Fundamental Research Funds for the Central Universities. F.G.X is supported in part by the National Program on Key R\&D Project of China (grant 2016YFA0400804), the Youth Innovation Promotion Association of CAS (CAS; id. 2016243), and the Natural Science Foundation of Shanghai (grant 17ZR1435800). The numerical calculations in this paper have been done on the supercomputing system in the Supercomputing Center of Wuhan University.

\end{document}